\documentclass[prl,twocolumn,showpacs,showkeys,preprintnumbers,amsmath,amssymb,superscriptaddress,aps,10pt]{revtex4-1}
\usepackage{graphicx}% Include figure files
\usepackage{dcolumn}% Align table columns on decimal point
\usepackage{bm}% bold math
\usepackage{natbib}
\usepackage{amsmath}
\usepackage{color}
\usepackage{ulem}

\begin{document}
%\preprint{preprint - not for distribution}
%
\title[PSL on NbSe2]{Air tightness of hBN encapsulation and its impact on Raman spectroscopy of van der Waals materials}
%Environmental effects in Raman measurements on NbSe$_2$ few-layer crystals
%Air tightness of different hBN encapsulation methods
%Air tightness of hBN encapsulation on SiO2 substrate. and impact on Raman spectroscopy of van der Waals materials 

\author{Johannes Holler}
\thanks{These two authors contributed equally to this work.}
\author{Lorenz Bauriedl}
\thanks{These two authors contributed equally to this work.}
\affiliation{Institut f\"ur Experimentelle und Angewandte Physik, University of Regensburg, 93053 Regensburg, Germany}
\author{Tobias Korn}
\affiliation{Institut f\"ur Physik, Universit\"at Rostock, 18059 Rostock, Germany}
\author{Andrea Seitz}
\affiliation{Institut f\"ur Experimentelle und Angewandte Physik, University of Regensburg, 93053 Regensburg, Germany}
\author{Furkan \"Ozyigit}
\author{Michaela Eichinger}
\author{Christian Sch\"uller}
\affiliation{Institut f\"ur Experimentelle und Angewandte Physik, University of Regensburg, 93053 Regensburg, Germany}
\author{Kenji Watanabe} \affiliation{National Institute for Materials Science, 1-1 Namiki, Tsukuba 305-0044, Japan}
\author{Takashi Taniguchi} \affiliation{National Institute for Materials Science, 1-1 Namiki, Tsukuba 305-0044, Japan}
\author{Christoph Strunk}
\author{Nicola Paradiso}\email{nicola.paradiso@physik.uni-regensburg.de}
 \affiliation{Institut f\"ur Experimentelle und Angewandte Physik, University of Regensburg, 93053 Regensburg, Germany}

\begin{abstract}

Raman spectroscopy is a precious tool for the characterization of van der Waals materials, e.g.~for the determination of the layer number in thin exfoliated flakes. For sensitive materials, however, this method can be dramatically invasive. In particular, the light intensity required to obtain a significant Raman signal is sufficient to immediately photo-oxidize few-layer thick metallic van der Waals materials.
In this work we investigated the impact of the environment on Raman characterization of thin NbSe$_2$ crystals. We show that in ambient conditions the flake is locally oxidized  even for very low illumination intensity. On the other hand, we observe no degradation if the Raman measurements are performed either in vacuum or on fully hBN-encapsulated samples. Interestingly, we find that covering samples deposited on the usual SiO$_2$ surface only from the top is not sufficient to prevent diffusion of oxygen underneath the layers.

\end{abstract}
\keywords{Raman spectroscopy, sensitive 2D materials, photo-oxidation, hBN encapsulation}
%\pacs{78.67.Ch, 78.30.-j, 63.22.Gh, 61.48.De, 52.40.Fd, 42.79.Dj}
%
\maketitle

2D materials are by definition \textit{all-surface} crystals, where the environment plays a crucial role compared to bulk systems~\cite{NovoselovReview2016}. Interaction with the substrate (typically a polished SiO$_2$ wafer) and absorption of ambient contaminants unavoidably introduce a considerable amount of disorder~\cite{Geim2013,QLi2019}. Further contamination is produced by standard exfoliation techniques based on viscoelastic polymers and by nano-lithographic patterning~\cite{Castellanos2014}. The impact of the environment is particularly critical for materials that tend to oxidize~\cite{QLi2019,Wood2014,YeSmall2016,Sun2017}. In this case, a brief exposure to air might lead to the complete disappearance of the material under study, which is rapidly substituted by its oxide species.

In most cases, oxidation in van der Waals materials is highly stimulated by light~\cite{Favron2015}. This makes their optical characterization, e.g.~by Raman spectroscopy, where relatively high illumination intensities and/or integration times are required, rather difficult.  On the other hand, Raman spectroscopy is a fundamental tool for the study of few-layer van der Waals materials. It provides a handy, precise and usually nondestructive means of mapping individual crystals. A plethora of important information, such as the number of layers~\cite{Ferrari2006,Heinz_ACSNano10}, doping~\cite{Pinczuk2007,Schueller2010,Sood2012} or applied strain~\cite{Huang2010,Novoselov13} can be extracted from Raman spectra for many different two-dimensional materials. While most Raman-active modes are specific to the crystal structure of a certain material, the rigid-layer shear and compression modes are generic to the layered van der Waals materials~\cite{Tan2012,Plechinger2012,Zhang2013,Nagler2016}. Due to the weak interlayer coupling in these materials, however, they have low Raman shifts and are more challenging to observe than higher-frequency modes.

%Raman spectroscopy is a key experimental tool for the investigation of the so-called van der Waals materials. Its range of applications goes from complex and delicate experiments to rapid characterization of the exfoliated flakes. An example of the latter case is the determination of the layer number in atomically thin crystals: the inter-layer shear mode in this case is strongly dependent on the total number of layers, and disappears altogether in the monolayer case. 

%Most van der Waals crystals (as, e.g., graphene, hBN and most semiconducting transition metal dichalcogenides)  are relatively inert and can be optically investigated without the need of particular precautions to avoid oxidation  or, in general, chemical degradation. 
%In this materials optical measurements can be carried out without particular limitations on the incident light intensity. On the other hand, many interesting materials are indeed extremely sensitive to light: in ambient conditions they photo-oxidize already under weak illumination intensity, well below the threshold for a significant Raman signal. 

In the present work we investigate the role of the environment in Raman spectroscopy of few-layer NbSe$_2$ crystals, an exemplary van der Waals material that is well known for its reactivity in air~\cite{ElBana2013,Sun2017,Cao2015}. 
We show that in ambient conditions NbSe$_2$ quickly oxidizes already under very low illumination intensity. The main goal of our work is to demonstrate convenient methods for preventing oxidation. We show that an encapsulation in hBN from both the bottom and the top side (here indicated as \textit{full} encapsulation) is totally air-tight and very effective in preventing photo-oxidation on a long term. Instead,  the commonly used \textit{half} encapsulation method (where NbSe$_2$ is covered by hBN only from the top side while adhering on the SiO$_2$ substrate) is not sufficient to protect the sensitive crystal, since it allows for a slow diffusion of oxygen underneath the layers.

Few-layer NbSe$_2$ crystals are exfoliated in a glove-box filled with nitrogen gas. Except for the glove-box, our stamping setup is nearly identical to that described in Ref.~\cite{Castellanos2014}. We use two methods for stamping 2D materials. The former method (hereafter indicated as \textit{simple stamping}) is the same as the one described in Ref.~\cite{Castellanos2014}: 2D crystals are exfoliated with an adhesive tape and transferred onto a thin polydimethylsiloxane (PDMS) film which adheres on a glass slide. Suitable flakes are found by inspection in an optical microscope and transferred onto a substrate using a micromanipulator placed under a zoom-lens. Our standard substrates consist of degenerately doped Si wafers capped with 285~nm-thick thermal SiO$_2$. The latter stamping method (hereafter indicated as \textit{pick-up}) is similar to that described in Ref.~\cite{Wangpickup,ZomerPC}:  2D crystals are first placed on temporary substrates (e.g.~by simple stamping) and then picked-up at 120~$^{\circ}$C by a thick, round PDMS film covered by polycarbonate (PC). Several different crystals can be sequentially picked up. Finally, the stack is stamped at 180~$^{\circ}$C onto the final substrate. At this temperature PC is dissolved and is released together with the stack onto the substrate surface. The PC residuals are finally dissolved in chloroform.

Raman spectroscopy measurements were performed in a self-built optical setup, experimental details are published elsewhere~\cite{Plechinger_2DMat, Nagler2016}. Briefly, we utilize a continuous-wave laser source with wavelength 532~nm filtered with a narrow bandpass. This is coupled into a 100$\times$, NA 0.8 microscope objective, resulting in a spot size of about 1~$\mu$m. The sample, mounted inside a small He-flow cryostat serving as a vacuum chamber, is positioned under the fixed optical beam path with a motorized xy-stage.
The backscattered Raman light is collected using the same objective, dispersed using a grating spectrometer equipped with a 1800 grooves per mm holographic grating and detected with a Peltier-cooled CCD camera. To ensure a wavenumber accuracy of 0.5~cm$^{-1}$, the instrument was periodically calibrated using the 520.5~cm$^{-1}$ band of the underlying Si substrate as a reference. While the higher-energy Raman modes of NbSe$_2$ could be measured using a single long-pass-filter and parallel polarization geometry, we had to utilize a set of 3 Bragg filters in order to access the low-frequency shear modes. Additionally, crossed-polarization geometry was used to further suppress the elastically backscattered laser.

\begin{figure}[tb]
\includegraphics[width=\columnwidth]{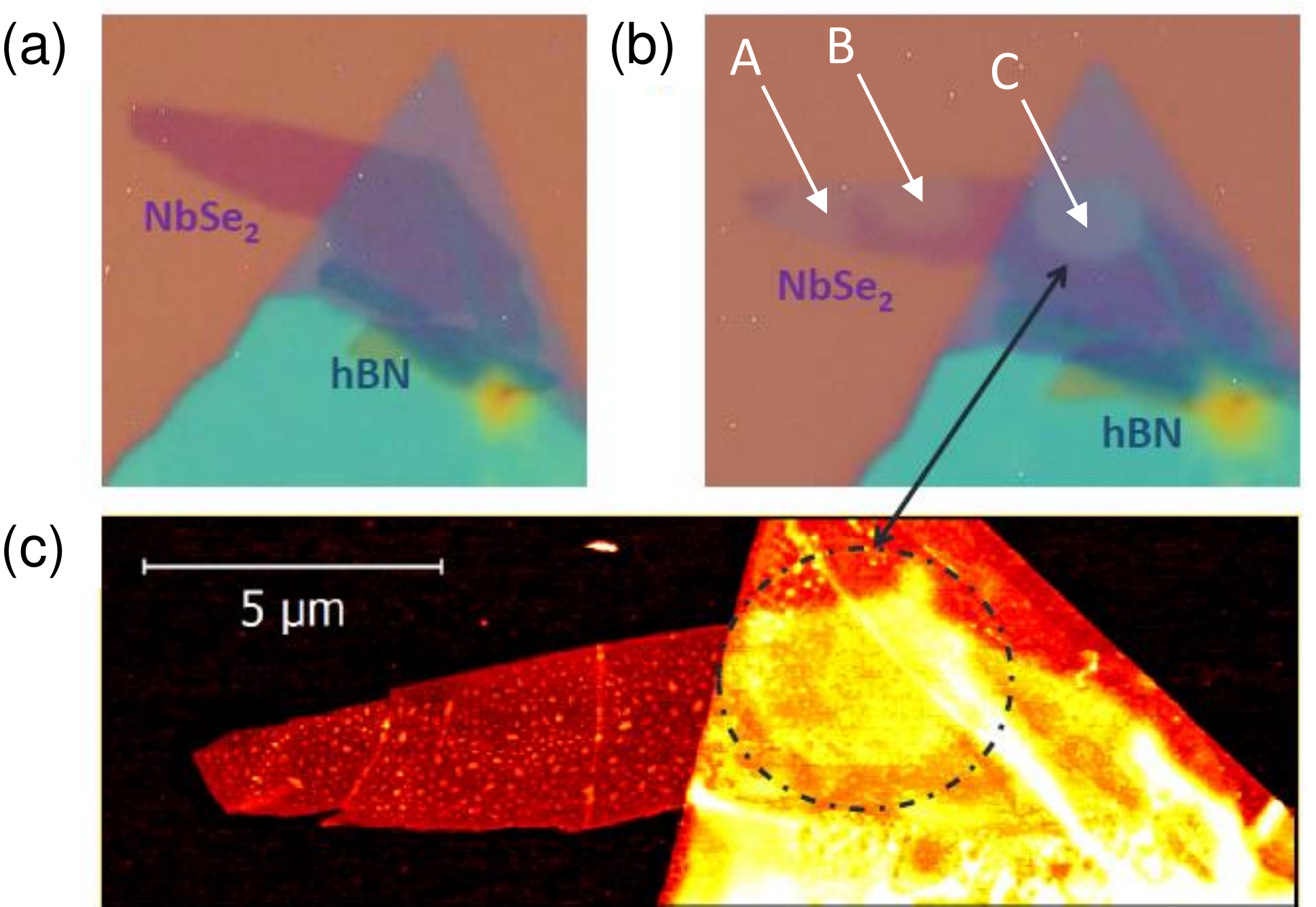}
\caption{(a) Optical microscope picture of a bilayer NbSe$_2$ flake (sample $\mathsf{A}$) partially covered by a hBN flake. (b) Optical picture of the same flake after the attempt of measuring the Raman spectra in three different positions, labelled as A, B and C in the figure. Owing to light-induced oxidation, these three spots appear as holes in the crystals. The incident light intensity equals 8~$\mu$W distributed over a  spot size of about 1~$\mu$m. The positions A and B correspond to unprotected NbSe$_2$ while the position C to an half encapsulated portion. (c) Atomic force microscopy topography of the same flake \textit{after} the measurements above. We notice that the oxidation does not produce an appreciable change in the apparent layer thickness for the spots A and B. The spot C in the encapsulated portion reveals instead a slight swell.  
}
\label{fig:burn}
\end{figure}

In our first reference experiment we investigated  few-layer NbSe$_2$ crystals transferred by simple exfoliation on a standard Si/SiO$_2$ substrate one day before the measurement. The results of micro-Raman measurements in ambient conditions 
show a dramatic oxidation within a few seconds even under an illumination power of only 8~$\mu$W.
%are spectacularly deceiving: even under an illumination power of only 8~$\mu$W, the NbSe$_2$ oxidizes within a few seconds. 
Figure~\ref{fig:burn}(a) shows an optical picture of sample  $\mathsf{A}$, an exemplary bilayer NbSe$_2$ flake partially covered by a hBN crystal. This sample has been obtained by the simple stamping procedure described above. Panel (b) shows the same sample after three brief (10~s integration each) attempts to measure the Raman spectrum in different positions, here labelled as A, B and C. The illumination power in this case is 8~$\mu$W, which is much less than the threshold for obtaining a discernible signal in reasonable integration times. After illumination, the spot positions appear as  holes, i.e.~regions that look more transparent than the surroundings. These are clear signs of photo-induced oxidation. Remarkably, the hBN protection does not help to limit the oxidation process: the spot C, which is covered by hBN, looks similar to those in A and B. This is one of the most surprising results of our work. Figure~\ref{fig:burn}(c) shows a topography map obtained by atomic force microscopy (AFM) on the same sample after the Raman measurement attempts. We notice that the oxidized regions are not visible on the unprotected part. This indicates that the oxidation does not significantly alter the thickness of the crystal (within the limit of vertical resolution of the AFM scan, which is about 0.5~nm). Interestingly, on the hBN-covered part, the oxidation spot appears instead as a bulge in the topography.

\begin{figure*}[tb]
\includegraphics[width=2\columnwidth]{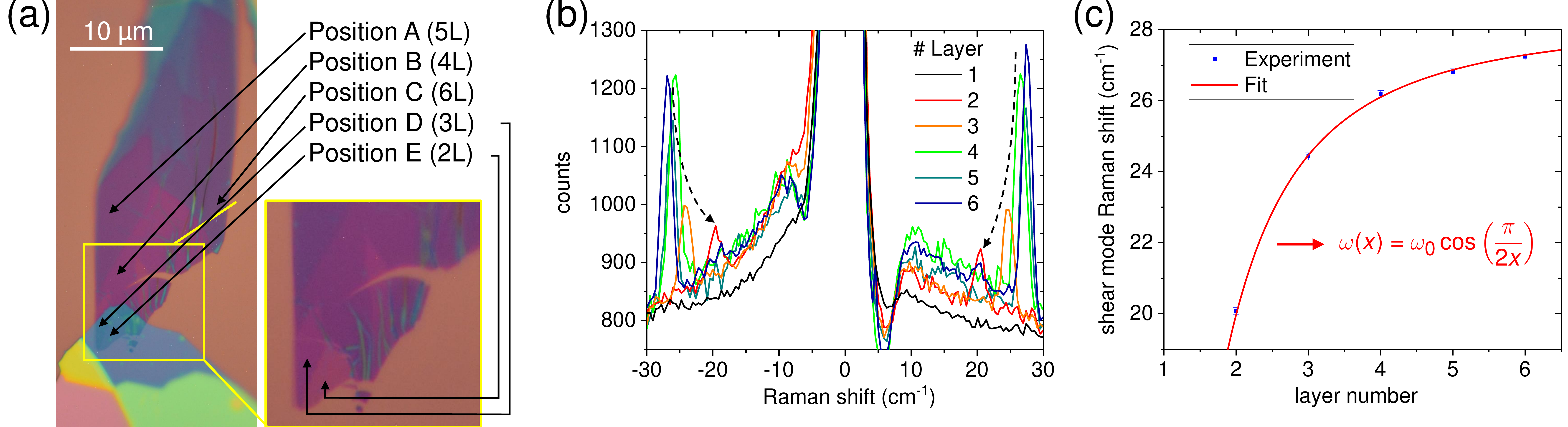}
\caption{(a) Optical microscope picture of sample $\mathsf{B}$. It consists of a NbSe$_2$ flake with many terraces, whose bottom part is covered by an hBN flake. Raman spectra have been measured on the positions A--E indicated by the arrows. The number of layer is indicated in brackets. The smaller panel shows a picture of the region corresponding to the yellow frame captured before stamping hBN.(b) Raman spectra measured on the positions A--E, plus a control measurement  (black curve) performed on a monolayer NbSe$_2$ flake on a different sample (sample $\mathsf{C}$, shown in the Supplementary Information). For all but the monolayer case, the Stokes and anti-Stokes $E^2_{2g}$ shear mode peaks are clearly visible. The mode frequency decreases with the number of layers, as indicated by the dashed arrows. (c) Plot of the Raman shift of the shear mode as a function of the layer number (blue dots) and corresponding fit with the  continuous model formula $\omega_s=\omega_0\cos (\pi/2N)$ (see text). From the fit we obtain $\omega_0=28.25\pm 0.10$~cm$^{-1}$.     
}
\label{fig:vacuum}
\end{figure*}

The most straightforward way to circumvent the oxidation problem is to perform the Raman measurements in high vacuum. This is the method used in the literature of Raman experiments on few-layer NbSe$_2$ crystals~\cite{Xi2015CDW,He2016}. We performed measurements on crystals obtained again by simple exfoliation on a standard Si/SiO$_2$ substrate one day before the measurement. This time, we loaded the samples in the vacuum chamber,  which was pumped down to few 10$^{-5}$~mbar. The measurements are carried out at room temperature. In such conditions, no oxidation is observed even for long exposures (of at least several tens of minutes) to 5~mW of illumination power, nearly three orders of magnitude more than the power used in the tests in air. The absence of oxidation allows us to use a suitable integration time so that we can also observe  weak Raman features such as the interlayer $E^2_{2g}$ shear mode~\cite{WangRaman1974,Pereira1982}. Our goal is to obtain a layer number mapping of the sample, in order to associate a well defined layer number to a given optical contrast. To this end, we chose a flake with several terraces, each with a different color and contrast. Part of the flake is again covered by hBN.  Figure~\ref{fig:vacuum}(b) shows low-frequency Raman spectra measured on five distinct positions on sample  $\mathsf{B}$ plus one spectrum measured on a monolayer portion located on  another sample (sample  $\mathsf{C}$). Figure~\ref{fig:vacuum}(c) shows a plot of the Raman shift versus number of layers, together with a fit with the function 
\begin{equation} 
\omega_S(N)=\omega_0\cos(\pi/2N).
\label{eq:eigen}
\end{equation} This expression is the result of a very simple model which assumes the layers to be identical objects of mass $m$, each connected by the same spring constant to the nearest neighbors. The eigenmodes of this classical mechanics problem~\cite{ThorntonMarion} are $$\omega(n,N)=\omega_0\cos(n\pi/2N),$$ where $n$ is the mode index. For the shear mode $E^2_{2g}$, the oscillation measured in Raman spectra corresponds to the highest frequency mode (i.e., $n=1$)~\cite{Lui2014,He2016}, which leads to the expression in Eq.\ref{eq:eigen}. 

\begin{figure}[t!]
\includegraphics[width=0.9\columnwidth]{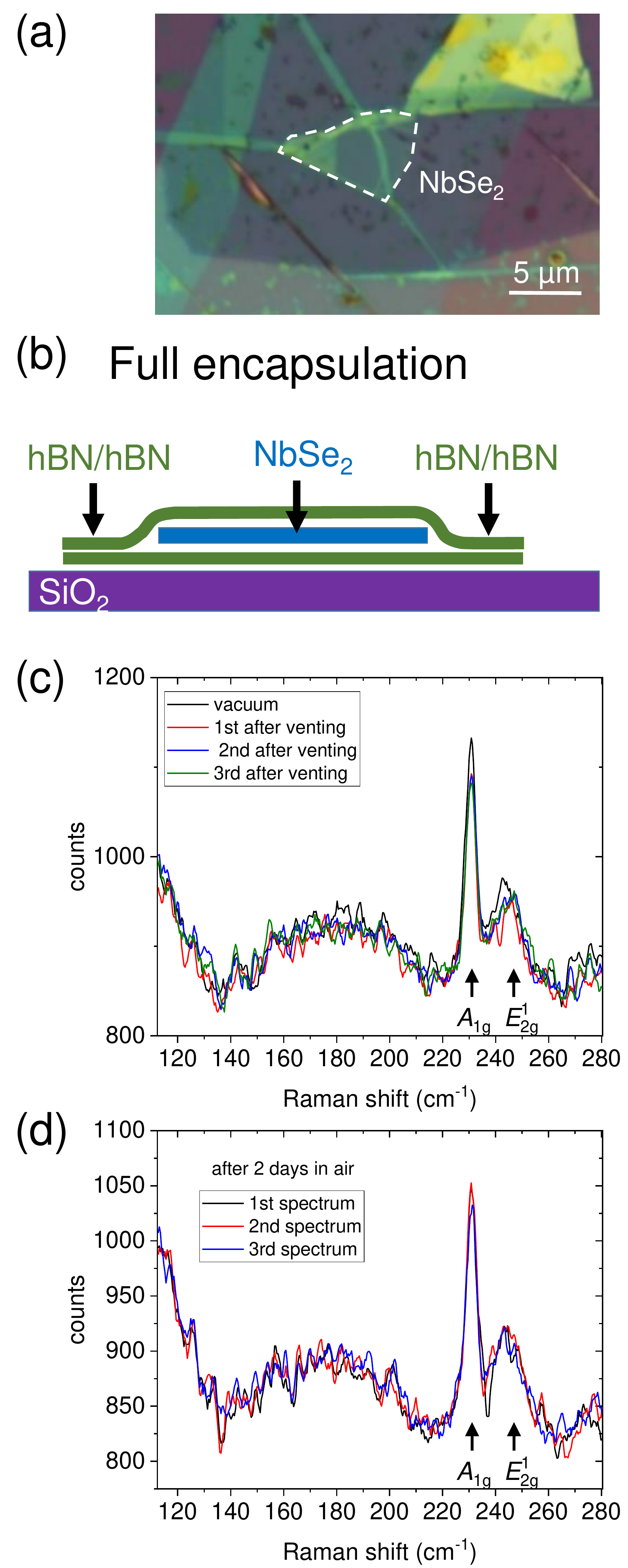}
\caption{(a) Optical picture of sample  $\mathsf{D}$, a few-layer NbSe$_2$ flake (indicated by a dashed line) fully encapsulated between two hBN flakes. (b) Sketch of the van der Waals stack. (c) Raman spectra in the range of the intralayer $A_{1g}$ and $E^1_{2g}$ modes. The spectra are measured in vacuum (black curve) and immediately after venting the chamber. (d) Series of Raman spectra measured after having exposed the sample to air for two days.
%The crystal portion overlapping contact \textit{4}, \textit{5} and \textit{6} is three layer-thick. (b) Color plot of the differential conductance between contact \textit{1} and \textit{5} as a function of voltage bias and temperature. 
%(c) Measured differential conductance between the same contacts, plotted as a function of current for $T=3.3$~K (orange line) together with the values calculated using the Skocpol-Beasley-Tinkham model with the same parameters as in Fig.~\ref{fig:simulcolplot}(b) (see text). The bias range corresponds to the orange-white dashed line in panel (b).  
%Finally, the magenta line shows the normalized BCS gap as a function of $t\equiv T/T_c$, i.e.~$\Delta(t)/\Delta(0)=\tanh \left(1.74\sqrt{t^{-1}-1}\right)$.
}
\label{fig:encaps}
\end{figure}

Despite the simplicity of the classical model,  its agreement with the experiment is remarkable. From the one-parameter fit we obtain the constant $\omega_0=28.25\pm0.10$~cm$^{-1}$. This value agrees well with the results of other measurements reported in the literature~\cite{He2016,Xi2015CDW}. We stress that, by definition, monolayers do not support interlayer modes, as observed in the measurement (black curve in Fig.~\ref{fig:vacuum}(b)).

%To identify the origin of the new set of Raman peaks, we have calculated the frequencies of all ZO′N(n) modes in FLG based on a coupled-oscillator model with only nearest-layer coupling. In this simple model, the layers are treated as a linear chain of N masses connected by constant springs. The predicted frequency of the ZO′N(n) normal modes can be expressed in a simple analytical form
%Useful References Nano Lett.2014 14 8 4615-4621,  Temperature-Activated Layer-Breathing Vibrations in Few-Layer Graphene
%a classical mechanics reference is "Classical Dynamics of Particles and Systems,  Brooks"

From the measurements on sample  $\mathsf{B}$ we also deduce that hBN encapsulation does not significantly affect the shear mode frequency. In fact, the shear mode Raman shifts measured on position D and E (trilayer and bilayer, respectively, see Fig.~\ref{fig:vacuum}(a)) are located on the same curve as the other points measured on position A, B, and C, see Fig.~\ref{fig:vacuum}(c). As discussed in the Supplementary Information, we have performed several control measurements with partially hBN-protected NbSe$_2$ flakes. The difference between the Raman shift on bare and protected NbSe$_2$ portions are always below the experimental uncertainty. This indicates that the differences in interlayer interaction and in layer density mechanically decouples  the TMDC from the hBN.

%With its large band gap hBN is an ideal protecting layer for optical experiments. However it is \textit{a priori} not obvious whether the hBN shear mode oscillations couple or not with those of the material under investigation. For NbSe$_2$ we found that this is not the case. The graph in Fig.~\ref{fig:vacuum}(c) clearly show that the data points corresponding to 2 or 3 layers are exactly where they are expected to be, without appreciable shift. 

So far, we have shown that performing the experiment in vacuum is the simplest way to perform a Raman measurement on sensitive van der Waals materials. However, this might be not always possible. 
%a clear example is a setup for tip-enhanced Raman spectroscopy (TERS), which is difficult to adapt to a vacuum chamber. More generally, a given setup or sample might not be compatible with measurements in vacuum. 
The main goal of our work is to seek a method to reliably protect a van der Waals flake from oxidation in situations where it is not possible to operate in vacuum. The method we found is the same used to obtain high-mobility graphene devices, namely, a full encapsulation in hBN~\cite{Wangpickup}. In  graphene devices, the main role of the bottom layer is to keep the flake spatially separated from the SiO$_2$ substrate surface, whose roughness and charge traps are highly detrimental for the mobility of the electron gas. In the present work, instead, the role of the bottom layer is to tightly seal the flake in between, due to the strong interaction between two adjacent hBN layers in the region outside, see sketch in Fig.~\ref{fig:encaps}(b). 

%One might expect that a single hBN layer on top of the flake might work as well. However, as shown below, we found that a single hBN layer on a standard SiO$_2$ substrate is not perfectly air tight. 

From a practical point of view, one could obtain a fully encapsulated NbSe$_2$ crystal using the simple stamping method described above. We empirically found, however, that this method does not work for NbSe$_2$. This latter does not reliably stick on hBN when transferred at room temperature using PDMS only. Fortunately, the pick-up method is instead much more reliable, probably owing to the squeezing of the inter-flake bubbles at the higher temperature required for the pick-up stamping~\cite{Wangpickup,ZomerPC}.

Figure~\ref{fig:encaps}(a) shows an optical picture of a NbSe$_2$ flake fully encapsulated in hBN, fabricated with the pick-up method. The key advantage of this configuration is that on the areas surrounding the NbSe$_2$ the top hBN flake adheres on another hBN flake. It is known from the high mobility graphene technology that such layers are in tight (on the atomic scale) contact and therefore the stack is expected to be air-tight.  We started our measurement session in vacuum. We  then measured  the Raman spectrum in the region around the $A_{1g}$ and $E^1_{2g}$ intralayer Raman modes~\cite{WangRaman1974,Pereira1982}. This Raman spectrum, measured with an illumination power of 2.6~mW and an integration time of 300~s, is shown in Fig.~\ref{fig:encaps}(c), black line. We then vented the system in air. We refocused the laser spot and immediately repeated the same measurement three times. We notice that, apart from a barely discernible decrease between the measurement in vacuum and the first measurement in air (which we attribute to a not perfect refocusing after the venting of the chamber), there is no significant change in the peak amplitude. This indicates that there is no appreciable degradation of the crystal induced by the light beam. To test the tightness of the stack over longer time scales, we left the system exposed to air for two days and then repeated the measurements (with illumination power 2.5~mW), which are shown in Fig.~\ref{fig:encaps}(d). Again, no degradation of the Raman signal was observed.

Let us sum up the results of our observations so far: optical measurements on few-layer NbSe$_2$ lead to a quick photo-induced oxidation within seconds even under an illumination of few $\mu$W/$\mu$m$^2$. In vacuum (pressure of the order of 10$^{-5}$~mbar) no signs of degradation are observed even for an illumination intensity of several mW/$\mu$m$^2$. A full encapsulation in hBN allows one to operate as if the sample were in vacuum, with no appreciable degradation under the same conditions. Finally, half encapsulation in hBN seems  unable to prevent oxidation in samples kept in air for days.

The results on half-encapsulated samples are the most interesting, since this method is the most commonly used in the literature to fabricate few-layer NbSe$_2$-based devices for transport measurements~\cite{Xi2015Ising,Xi2016Gate,ParadisoPSL2019}. In fact, a common way to obtain devices for 4-terminal measurements consists in stamping the exfoliated flakes directly on pre-patterned Au electrodes and then stamping an hBN flake on top as a protection against oxidation. The question is then whether the hBN protection is indeed air-tight: our measurements above seem to indicate that this is not the case. However, it is important to determine whether the photo-induced oxidation observed, e.g., on sample  $\mathsf{A}$ is due to the SiO$_2$ substrate or to oxygen molecules diffusing in between the layers from the environment.   
	
\begin{figure}[t!]
\includegraphics[width=0.9\columnwidth]{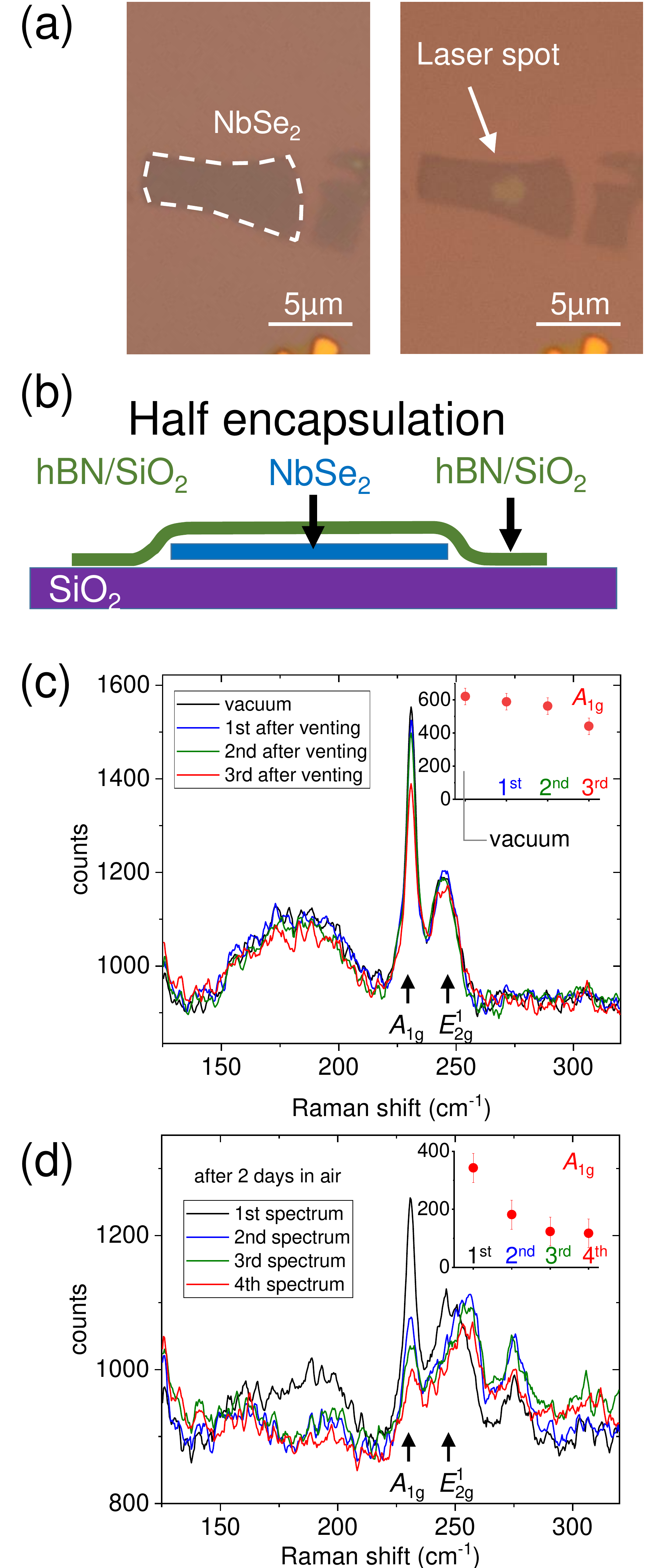}
\caption{(a) Optical pictures of sample  $\mathsf{E}$, a half hBN encapsulated few-layer NbSe$_2$ flake, taken before (left) and after (right) the Raman measurements. The edges of the top hBN flake lie outside the picture frame.   (b) Sketch of the van der Waals stack. (c) Raman spectra in the range of the intralayer $A_{1g}$ and $E^1_{2g}$ modes, measured in vacuum (black curve) and immediately after venting the chamber. (d) Series of Raman spectra measured (as for sample  $\mathsf{D}$) after having exposed the sample to air for two days.
%(c) Measured differential conductance between the same contacts, plotted as a function of current for $T=3.3$~K (orange line) together with the values calculated using the Skocpol-Beasley-Tinkham model with the same parameters as in Fig.~\ref{fig:simulcolplot}(b) (see text). The bias range corresponds to the orange-white dashed line in panel (b).  
%Finally, the magenta line shows the normalized BCS gap as a function of $t\equiv T/T_c$, i.e.~$\Delta(t)/\Delta(0)=\tanh \left(1.74\sqrt{t^{-1}-1}\right)$.
}
\label{fig:halfencaps}
\end{figure}

%The last result is the most puzzling, since it raises the question whether the oxidation results exclusively from the contact with the SiO$_2$ substrate, or whether it is the half encapsulation that is not air tight, so that it enables oxygen molecules to slowly diffuse in between the layers. 
A direct way to test this is to compare the Raman spectra measured on the same half encapsulated sample (i) in vacuum, (ii) immediately after venting in air, and (iii) after a given time of exposure to air, in the same way as done for sample $\mathsf{D}$. The results of such test are shown in Fig.~\ref{fig:halfencaps}. Here we study a NbSe$_2$ flake (sample $\mathsf{E}$) exfoliated in N$_2$ atmosphere and half encapsulated in hBN before being removed from the glove box. The sample is mounted in the vacuum chamber and Raman spectra are measured at a pressure of few $10^{-5}$~mbar (panel b), with an illumination power of 2.5~mW and 300~s integration time. The chamber is vented with air and then several spectra are acquired in the next tens of minutes. \textit{In this case, no appreciable degradation is observed} (panel (c)), except for a small decrease in the peak amplitude observed in the last spectrum, after more than 10 minutes of total illumination.  The sample is then left in air for two days. Then, several Raman spectra are measured (panel (d)). Now, every measurement (300~s integration time with a power of 2.3~mW) produces a substantial decrease of the Raman peak amplitude, until only a faint trace is left after ten minutes of exposure. Similar results have been obtained on another half encapsulated sample, as discussed in the Supplementary Information.

These results clearly indicate that \textit{half hBN encapsulation is not air tight}. The subnanometric roughness of the SiO$_2$ surface allows oxygen molecules to slowly diffuse in between the interface SiO$_2$-hBN. On the other hand,  the atomically smooth hBN-hBN interface is perfectly tight and ensures a long term and reliable protection against oxidation.

This behavior seems to be confirmed also by our experience with the contact resistance of Au or graphite contacts on NbSe$_2$ in half encapsulated devices. We systematically observe that the contact resistance monotonically increases with the exposure time to air. Based on the above results, we suggest to use for transport experiments either a full encapsulation in hBN (as in Ref.~\cite{Tsen2015}) or, if this is not possible or convenient, we recommend to keep the sample as much as possible in high vacuum and minimize the exposure time to air.

%A third sample with no hBN coverage whatsoever was also loaded in the chamber in the same measurement session. There we observe an immediate degradation of the signal, as discussed in the Supplementary Information.     

%During the same measurement session we investigated also a half encapsulated crystal. The idea here is to understand the time scale required to oxygen to diffuse within the layers. The results of this measurements are shown in \textbf{show the flake and the progression of measurements soon after venting and after two days.}

In conclusion, we have investigated how exposure to air affects Raman measurements on NbSe$_2$, a representative van der Waals material which tends to oxidize in the presence of air and intense illumination.  We found that samples in high vacuum are not affected by oxidation even under illumination intensities of several mW/$\mu$m$^2$. We demonstrated that full encapsulation in hBN effectively provides a long term protection against oxidation for samples kept in ambient conditions. The same does not hold for half encapsulated devices, where we observed a slow diffusion of oxygen in between the hBN layer and the SiO$_2$ substrate.

\section*{Supporting Information Available}
(Absence of) influence of hBN encapsulation on the shear mode Raman shift. Measurements on monolayers: sample $\mathsf{C}$. Additional measurements on half hBN encapsulated samples.
\section*{ACKNOWLEDGEMENTS}
\begin{acknowledgments}
The work was funded by the Deutsche Forschungsgemeinschaft within Grants
DFG SFB1277 (B04 and B06), GRK1570 and KO3612/3-1. Bulk NbSe$_2$ was purchased from HQ Graphene.
\end{acknowledgments}
\bibliographystyle{achemso}
\bibliography{biblio}
\end{document}